\documentstyle[prl,twocolumn,aps,psfig,floats]{revtex}
\begin{document}                
\draft
\def\be{\begin{equation}}
\def\ee{\end{equation}}
\def\ba{\begin{eqnarray}}
\def\ea{\end{eqnarray}}
\title{Scars of Invariant Manifolds in Interacting Chaotic Few--Body Systems}
\author{T. Papenbrock$^{1}$, T. H. Seligman$^2$ and H. A. Weidenm\"uller$^1$}
\address{$^1$Max--Planck--Institut f\"ur Kernphysik,
Postfach 103980, 69029 Heidelberg, F. R. of Germany}
\address{$^2$University of Mexico (UNAM), Instituto de Matem\'aticas,
Unidad de Cuernavaca, 62191 Cuernavaca, Mexico}
\maketitle
\begin{abstract}
We present a novel extension of the concept of scars for the wave
functions of classically chaotic few--body systems of identical
particles with rotation and permutation symmetry. Generically there
exist manifolds in classical phase space which are invariant under the
action of a common subgroup of these two symmetries. Such manifolds
are associated with highly symmetric configurations. If sufficiently
stable, the quantum motion on such manifolds displays a notable
enhancement of the revival in the autocorrelation function which is
not directly associated with individual periodic orbits. Rather, it
indicates some degree of localization around an invariant manifold 
which has collective characteristics
that should be experimentally observable.

\end{abstract}
During the last decade, spectral fluctuation properties of classically
fully chaotic systems (``quantum chaos'' for short) have been largely
understood, and interest has shifted to the non--generic properties of
wave functions. In systems with few degrees of freedom, these
functions may display scars
\cite{McDoPhD,Heller,Bogomolny88,Berry89,Agam}, {\it i.e.} possess
increased intensities along short and not very unstable periodic
orbits of the corresponding classical system.  

In this Letter, we extend the study of scars to systems with identical
particles. Such systems possess symmetry properties which are
generically absent in systems with two or three degrees of freedom.
Particle identity leads to permutational symmetry. In addition,
a self--bound system possesses rotational symmetry. We show that the
combination of these two symmetries leads to a novel mechanism for the
formation of scars.

In a previous paper, two of the authors studied the influence of both
symmetries on the periodic--orbit structure of chaotic systems
\cite{PaSel}. In the present Letter, we procced differently. The
combination of rotational and permutational symmetry allows us to
construct invariant manifolds in classical phase space. We show that
such manifolds may lead to scarred wave functions of the corresponding
quantum system. The scars are not related to individual periodic
orbits but are similar to the ones found by Prosen in billiards
\cite{Prosen97}. We believe that our results are important for finite
many--body systems such as atoms, molecules or atomic nuclei where
this generalized scarring has obvious implications for collective
motion. Our results derive from the numerical study of a specific
Hamiltonian. For reasons given below, we believe our results to be
generic, however.

The paper is organized as follows. We define a rotationally invariant
few--body system of four interacting identical particles in two
dimensions. We prove the existence of an invariant manifold and
analyze its classical stability. Finally, we investigate the
corresponding quantum system. This is done by semiclassical
propagation of wave packets.
 
Our Hamiltonian has the form 
\be
\label{fullHam}
H = \sum_{i=1}^4\left(\frac{1}{2m}{\bf p}_i^2 
+ 16\alpha|{\bf r}_i|^4\right) 
- \alpha\!\!\sum_{1\le i < j\le 4}\!\!|{\bf r}_i - {\bf r}_j|^4,
\ee
where ${\bf p}_i\!=\!(p_{x_i},p_{y_i})$ and ${\bf r}_i\!=\!(x_i,y_i)$ with 
$i = 1, \ldots, 4$ are the two--dimensional momentum and position vectors of 
the $i^{th}$ particle, respectively. We use units where 
$m=\alpha=1, \hbar=0.01$; 
then coordinates and momenta are given in units of $\hbar^{1/3} \alpha^{-1/6}
m^{-1/6}$ and $\hbar^{2/3}\alpha^{1/6}m^{1/6}$, respectively. For the 
Hamiltonian, this leads to the scaling relation $H(\gamma^{\frac{1}{2}}{\bf p},
\gamma^{\frac{1}{4}}{\bf r})=\gamma H({\bf p},{\bf r})$. This shows that the 
structure of classical phase space is independent of energy. Moreover, energy 
and total angular momentum are the only integrals of motion, and the system is 
non--integrable.

\begin{figure}[htb]
  \begin{center}
      \leavevmode
    \parbox{0.50\textwidth}
           {\psfig{file=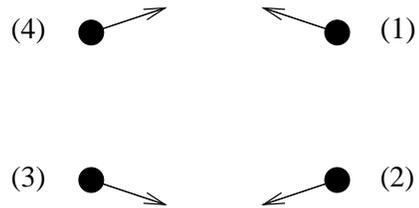,width=0.30\textwidth,angle=270}}
  \end{center}
  \caption{Collective configuration}
  \label{FigConf}
\end{figure}

{\it Construction of the invariant manifold.} The Hamiltonian (\ref{fullHam}) 
is invariant under the symmetric group $S_4$ and the orthogonal group $O(2)$. 
Thus, we may apply the ideas of ref. \cite{PaSel}. By imposing suitable 
symmetry requirements onto the initial conditions, we can restrict classical 
motion to a low--dimensional invariant manifold. Many such manifolds exist 
and it suffices to consider one of them. We choose the initial conditions in 
such a way that positions and momenta exhibit the symmetry of a rectangle.
Such a configuration is shown in Fig.~\ref{FigConf}. Positions and momenta of 
the particles are indicated by points and arrows, respectively. Obviously, 
particle 3 is the image of particle 1 under inversion, whereas particle 2 and 
particle 4 are the mirror images of particle 1 under a reflection at 
appropriately chosen axes, respectively. The manifold is spanned by the two 
two--dimensional vectors ${\bf p}=(p_x,p_y)$ and ${\bf r}=(x,y)$ giving the 
momentum and position of particle 1. The associated Hamiltonian 
$\tilde{H}({\bf p},{\bf r})=\frac{1}{2}{\bf p}^2 + 16x^2y^2$ has been studied 
extensively in the literature, both in the classical and the quantum case 
\cite{Simon83,Waterland89,Tomsovic91,Dahl91,BoToUl,Eck}. 
The classical system is essentially chaotic. Only one stable periodic
orbit is  known which is surrounded 
by a very small island of stability \cite{Dahl90}. The periodic orbits can be 
enumerated by means of an (incomplete) symbolic code consisting of three 
letters \cite{Dahl91}. Application of rotations and/or permutations to the 
configuration shown in Fig.~\ref{FigConf} yields further equivalent invariant 
manifolds.

{\it Stability of motion close to the invariant manifold.} For
particle positions far from the origin. we may expand the
potential of the Hamiltonian~(\ref{fullHam}) around the ``channel''
configuration defined by the position $x=r\gg E^{1/4}, \ y=0$ of
particle 1. This yields a quadratic form with non--negative
eigenvalues proportional to $r$. Thus, far out in the channel, the
particles perform almost harmonic oscillations with high frequencies
in the directions transverse to the channel while moving slowly in the
direction of the channel. This confines the motion to a small vicinity 
of the invariant manifold. Moreover, this motion is regular. Close to
the central region, the motion is chaotic. Hence, we expect our
few--body system to display intermittency.

To analyze the stability of the central region, we confine ourselves to 
computing the {\it full} phase--space monodromy matrices of certain periodic 
orbits within the manifold. All orbits up to code length three (12 in total) 
are taken into account. Together they explore a significant part of the 
central region. For each periodic orbit there are five pairs of stability 
exponents that correspond to the directions perpendicular to the invariant 
manifold. A sixth pair of stability exponents corresponds to the motion within 
the invariant manifold. The remaining two pairs of stability exponents vanish
because of the continuous symmetries of the Hamiltonian. Our computations show 
that the motion in the directions perpendicular to the manifold is unstable. 
However, the corresponding stability exponents are significantly smaller than 
the stability exponents which correspond to the chaotic motion within the 
invariant manifold, see Table~\ref{tabstab}. This combination of large 
stability exponents within the manifold and small stability exponents 
perpendicular to the manifold leads us to conjecture that the manifold may 
scar wave functions of the corresponding quantum system.

Clearly this is a particular property of the invariant manifold 
we have chosen. If we 
consider the collinear configuration with mirror symmetry about the 
center of the system, we obtain a different invariant manifold
 with very large transverse 
Lyapunov exponents. On this invariant manifold we do not expect scarring.

{\it Quantum computation.} To prove our conjecture 
we consider the time evolution of a Gaussian wave packet
\begin{equation}
\Psi({\bf r},t)=c\exp\left[-\frac{1}{2}({\bf r}-{\bf r}_0)^TA({\bf r}-
{\bf r}_0)+\frac{i}{\hbar}{\bf p}_0^T({\bf r}-{\bf r}_0)\right]
\end{equation}
where $<\bf p>= {\bf p}_0$ and $<\bf r>={\bf r}_0$ define a point on the
invariant manifold.
We have used the shorthand notation
${\bf r}=({\bf r}_1,{\bf r}_2,{\bf r}_3,{\bf r}_4)$ and
${\bf p}=({\bf p}_1,{\bf p}_2,{\bf p}_3,{\bf p}_4)$
for configuration and momentum space vectors, respectively.
The autocorrelation function $C(t)=<\Psi(t=0)|\Psi(t)>$ is computed
in semiclassical approximation. Within the manifold we used Heller's
cellular dynamics \cite{celldyn} which takes into account the
nonlinearity of the classical motion. In the transverse direction the
time--propagation was done using linearized dynamics only. This approximation 
neglects any recurrences from the transverse directions and implies a
permanent flux of probability out of the manifold and its vicinity. 
On the time scales considered here, the linearization is justified since
the classical return probability to the manifold of transversely escaping 
trajectories is negligible. It is also important to note that the loss
of probability inside the manifold is not severe since the transverse 
stability exponents are not too large. 

We launch wave packets along periodic or aperiodic orbits lying within
the invariant manifold and consider their revival as measured by the
autocorrelation function. To achieve shorter recurrence times the
initial packet was symmetrized with respect to the reflection symmetry
of the system within the invariant manifold. This effects a partial
projection onto the symmetric and eliminates the antisymmetric
representation of ${\cal S}_4$; antisymmetrizing with respect to the
reflections we could have eliminated the symmetric rather than the
antisymmetric representation.
 
\begin{figure}[htb]
  \begin{center}
      \leavevmode
    \parbox{0.50\textwidth}
           {\psfig{file=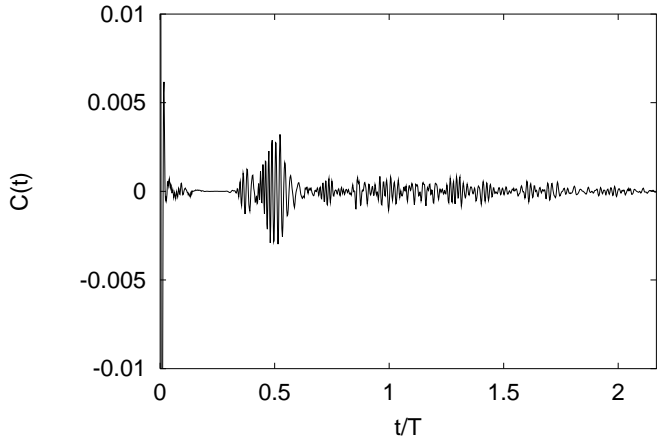,width=0.50\textwidth,angle=270}}
  \end{center}
  \caption{Autocorrelation function $C(t)$ of a symmetrized wave packet 
  launched on a periodic orbit with period $T$ inside the weakly unstable 
  manifold. In addition to the linear revival around $t=\frac{T}{2}$, a strong
  nonlinear revival is seen for larger times.} 
  \label{Fig2a}
\end{figure}

\begin{figure}[htb]
  \begin{center}
      \leavevmode
    \parbox{0.50\textwidth}
           {\psfig{file=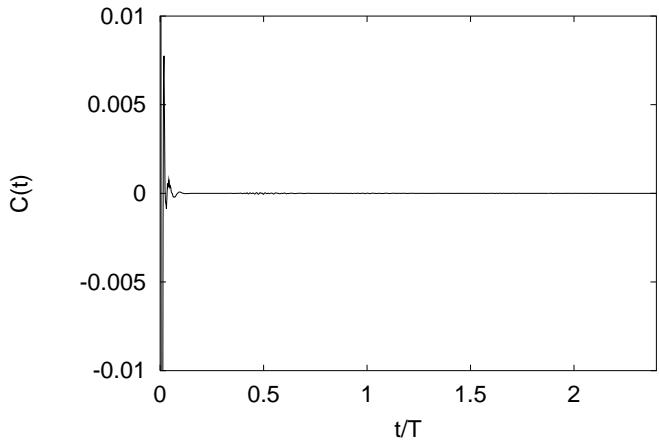,width=0.50\textwidth,angle=270}}
  \end{center}
  \caption{Autocorrelation function $C(t)$ of a symmetrized wave packet 
   launched on a short periodic orbit with period $T$ inside the very
   unstable collinear manifold. Practically no revival is seen.}
  \label{Fig2b}
\end{figure}

We propagate such wave packets for both, the manifold with small and
with large transversal exponents. For not too unstable periodic orbits
we expect a fairly strong revival after one period, known as the
linear revival \cite{celldyn}. As an example, we show in
Fig.~\ref{Fig2a} the real part of the autocorrelation function,
calculated for the more stable invariant manifold and starting on the
periodic orbit 2. We indeed find strong linear revival. However, at
larger times we find randomly scattered strong revivals, the revival
corresponding to the second period not being dominant. This implies
that a significant fraction of the original amplitude remains within
the invariant manifold, and that this fact is not related to the
periodic orbit we started on. Revivals calculated for packets started 
on aperiodic orbits show similar features except for the obvious 
absence of the linear revival. The collinear and more unstable of the 
two manifolds discussed above, on the contrary, shows practically no 
revival at all. Even the linear revival is negligible, see Fig.~\ref{Fig2b},
despite the fact that the packet was launched along a short periodic
orbit.

\begin{figure}[htb]
  \begin{center}
      \leavevmode
    \parbox{0.50\textwidth}
           {\psfig{file=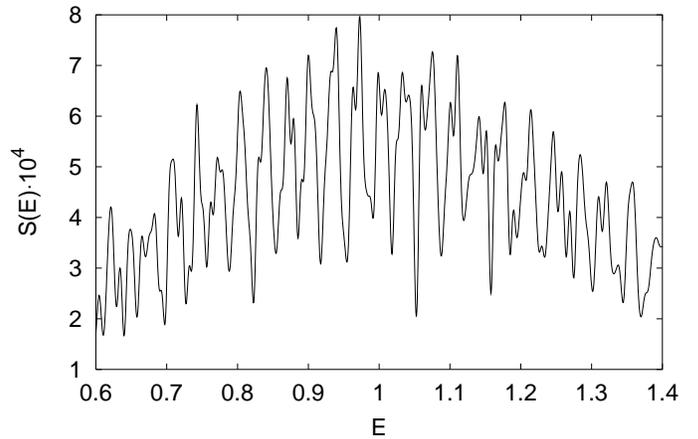,width=0.50\textwidth,angle=270}}
  \end{center}
  \caption{ Fourier transform $S(E)$ of the autocorrelation function $C(t)$
   shown in Fig.~\protect\ref{Fig2a}. The equidistant structure
   corresponding to half the period of the periodic orbit is accompanied
   by a fine structure resulting from the nonlinear revival.}
  \label{Fig3}
\end{figure}

Fig.~\ref{Fig3} shows the Fourier transform of the autocorrelation
function whose real part was displayed in Fig. \ref{Fig2a}. Our time
sequences are not long enough to resolve the spectrum, and it seems
difficult to obtain sequences of sufficient length  with a reasonable
numerical effort. Nevertheless the Fourier transform in
Fig.~\ref{Fig3} is very informative. It shows a peak in energy
(frequency) corresponding to the energy of the original wave
packet. We have chosen the initial conditions $({\bf p_0, r_0})$ such 
that the energy of the initial wavepacket is centered around $E=1$.
An integration over classical phase space shows that this energy is
chosen around the $10^{10 {\rm th}}$ level. This
implies that we are deep within the semiclassical region. The
expected peak in energy has a superposed equidistant structure with 
spacing $\Delta E= 0.034$ 
corresponding to the period of the orbit on which we launched the wave
packet. In addition there is a significant superposition of other
frequencies resulting in a finer, non-equidistant structure. 

This indicates that other trajectories
in the manifold contribute. The structure under the peak is typical
for the more stable invariant manifold and practically absent for the
collinear one (not shown).

Our argument suggests that the enhanced revival seen on sufficiently
stable invariant manifolds should not depend on the irreducible
representation of ${\cal S}_4$ we choose, unless the invariant manifold 
itself limits the possible representations. We could not check this
point since our calculation did not project onto a specific
irreducible representation. Nevertheless, the localization effect
implicit in the large revival of the autocorrelation function
is not of classical nature. Indeed, an ensemble of trajectories
launched near the invariant manifold did not return near it on the
time scale we consider. We note that this observation also justifies
the linear approximation for cellular dynamics in directions
transversal to the invariant manifold.

In summary, we have shown that wave packets may have unusually long
life times on certain invariant manifolds characterized by small
classical transverse instabilities. This is a quantum and not a
classical phenomenon and constitutes a novel and very exciting
extension of the concept of a scar to which we attribute considerable
significance. Indeed, invariant manifolds of the type considered above
occur generically in many--body systems of physical interest like
nuclei, atoms, molecules, or small metallic clusters. The importance
of these manifolds for quantum properties will hinge on their
stability properties: Stability will determine the degree to which
scars actually exist in such systems. Stability is a system--specific
property, of course, and generic statements are at least very
difficult. We recall, however, that in Helium the collinear manifold
\cite{WiRiTa} (which is linearly stable in the perpendicular direction) 
supports doubly excited states.  This shows why we expect our results to be
generic for interacting few--body systems.

\begin{table}[htb]
\begin{center}
\leavevmode
\begin{tabular}{|c||r|r|r|c|}
Code & $T$   & $u_\parallel$ &$\sum_{i=1}^nu^{(i)}_\perp$ & $n$ \\ \hline
 1   & 3.313 &  5.74         &  ---                       &  0  \\
 2   & 2.622 &  4.86         & 3.29                       &  2  \\
 01  & 2.958 &  4.46         & 1.22                       &  2  \\
 02  & 4.036 &  6.99         & 4.18                       &  3  \\
 12  & 7.933 & 11.67         &  ---                       &  0  \\
 001 & 8.015 &  9.64         &  ---                       &  0  \\
 002 & 5.074 &  8.07         & 6.84                       &  4  \\
 011 & 4.541 &  7.30         & 0.35                       &  2  \\
 012 & 5.753 & 10.00         & 1.63                       &  1  \\
 022 & 6.623 & 11.37         & 7.13                       &  5  \\
 112 & 2.860 &  4.79         & 0.68                       &  1  \\ 
 122 &13.259 & 23.03         &21.41                       &  5  \\   
\end{tabular}
\protect\caption{\protect\small
Periods and stability exponents of periodic orbits up to code 
length three lying inside the invariant manifold.  
$T$ is the period, $u_\parallel$ the stability
exponent inside the manifold,  $\sum_{i=1}^n u^{(i)}_\perp$ the
sum of all ($n$ in total) real positive stability exponents perpendicular to 
the manifold. The energy of the orbits is $E=4$.}
\label{tabstab}
\end{center}
\end{table}

\end{document}